# MAPbBr3 monocrystals under electron beam — radiolysis and degradation revealed by cathodoluminescence spectroscopy

**Authors:**

Yu. O. Kulanchikov[1,2], P.S. Vergeles[1], K. Konstantinova[2], A.R. Ishteev[2], D.S. Muratov[2], E.E. Yakimov[1], E.B. Yakimov[1,2] and D.S. Saranin[2*].

**Affiliations:**

[1]Institute of Microelectronics Technology RAS, Chernogolovka 142432, Russia

[2]Laboratory of Advanced Solar Energy, MISIS University, Moscow, Leninskiy pr.6, 119049, Russia

[3]CHOSE–Centre for Hybrid and Organic Solar Energy, University of Rome Tor Vergata, Via del Politecnico 1, 00133 Roma, Italy

*Corresponding author:

Dr. Danila S. Saranin saranin.ds@misis.ru

**Abstract**

Study of the local optical properties using electron beam (e-beam) can provide a valuable information concerning the inspection of the materials quality, the presence of the different phase inclusions and defects. Halide perovskites have been shown to be highly sensitive to external stress conditions like ambient atmosphere, light, and heat. In this paper, the cathodoluminescence (CL) spectroscopy has been exploited to carry out the investigation of $CH_3NH_3PbBr_3$ monocrystals under low energy electron beam irradiation. The CL spectra exhibited strong transformation with the increase of the irradiation dose and significant shifts of the peak maximums from 2.23 eV to >2.5 eV. Utilizing a larger e-beam energy (>20 keV) was found to be preferable to slow down the dynamics of the decomposition and corrosion. The mechanisms of the changes in MAPbBr3 properties after e-beam exposure and correlation to the in-depth distribution of deposited energy were discussed.

Hybrid halide perovskites (**HP**s) possess outstanding optical absorption in the visible range (~$10^5$ cm$^{-1}$),[1,2] high charge carrier lifetimes[3] and mobilities,[4,5] making them superior to photovoltaics[1] and ionizing radiation detection tasks.[7] The monocrystalline (MNC) HPs could be fabricated using the low-temperature solution growth,[8] which are cost-effective alternative to the standard expensive manufacturing of the semiconductors (Si, GaAs, Ge, GaN, etc.).[9] However, several instability factors of HP limit its exploitation in the devices. Thermal decomposition[2] of HPs occurs at relatively low temperatures with formation of the volatile products[3] (methyl-ammonium gas) and phase segregation[4] (lead-halides, etc.). The presence of moisture could trigger the oxidation process[5] at the surface of HP and initiate non-reversible corrosion[6]. Structural changes of the HP MNCs take place at the micro-scale; therefore, the special characterization by the transmission and scanning electron microscopy should be used. A proper assessment of the electron beam-induced damage on the materials is needed to provide the inspection at the micro-scale. The electron beam excitation can be used for both the characterization of local electrical and optical properties and monitoring their changes due to irradiation. Besides, such investigations allow to estimate the prospects of HPs application for charge particle detection. The effects of low energy electron beam irradiation (LEEBI)[7–9] as well as X-ray irradiation[10] on the properties of organic–inorganic hybrid perovskites are attracting a lot of scientific interest. The interaction of HPs with X-ray photons leads to breaking [11] the chemical bonds and molecular decomposition, accompanied by the release of metallic lead. The LEEBI is carried out in a vacuum and conditions of the low-pressure can accelerate the formation of the gas products (typically MA)[12] and evaporation from the degraded surface. Despite significant efforts, a comprehensive understanding of the mechanisms that drive degradation pathways in halide perovskites under ionizing irradiation has yet to be achieved.

The cathodoluminescence (CL) spectroscopy could be used as an effective characterization tool for HPs[13]. The beam of the accelerated electrons directed to the semiconductor can penetrate to the volume of the material and induce luminescence that provides information about the material's band structure, defect states, and charge carrier dynamics[14]. This method allows to monitor the optical properties to a depth exceeding 10 μm although it is also sensitive to the surface transformations, leading to a change of surface recombination velocity. The light emission spectra can be figured out with the variation in the material's depth from units to tens of microns used to study the depth dependence of optical properties [15]. The penetration capability of CL could be adjusted by varying the beam energy. This paper displays a comprehensive study of the optical properties of $CH_3NH_3PbBr_3$(MAPbBr$_3$) MNCs in relation to LEEBI. We performed cathodoluminescence measurements with different beam energy $E_b$ in the range from 2.5 to 30 keV and investigated the evolution of the spectra with e-beam irradiation doses. We observed the considerable "blue" shift of the CL with the increase of electron irradiation dose. The features of the depth dependencies on the e-beam energy were discussed.

The interaction of the e-beam with materials depends on several parameters–penetration depth, irradiation dose and beam current. In the current work, we used solution-frown MAPbBr$_3$ MNC (see experimental in the supplementary for the details, including XRD spectra and photo-images). First, we analyzed the in-depth distribution of deposited energy in $CH_3NH_3PbBr_3$ ($MAPbBr_3$) calculated by the Monte-Carlo method using in-house Monte Carlo program based on the algorithm described in [16] for the beam energies ($E_b$) in the range of 2.5–30 keV, which are shown in **Fig. 1 (a)**. The primary electron penetration depth (electron range R) can be obtained from such a distribution as a layer thickness, in which 99% of beam energy is deposited. As our calculations shown, R can be approximated as R = $0.013 \times E_b(keV)^{1.75}$ μm. The density of the deposited energy (dE/dz) decreases with a rise in $E_b$ (see Fig. 1) roughly as $E_b^{-0.75}$. Electron-hole pair generation rate can be obtained by a division of the deposited energy depth distributions by the electron-hole pair creation energy $E_i$, which, as shown in [17], can be approximated as $E_i = 2.8 E_g + 0.6 eV$ [18], where $E_g$ is the band-gap. We performed the measurements of the cathodoluminescence for "as fabricated" $MAPbBr_3$ MNC with 30 keV beam irradiation (**fig.1b**). The observed CL spectra of the $MAPbBr_3$ can be fitted with two Gaussian curves $\exp[-(E-E_m)^2/2\sigma^2]$ (see experimental for the details) with energies $E_m$ of 2.26 and 2.23 eV and σ equals to 30 and 60 meV.

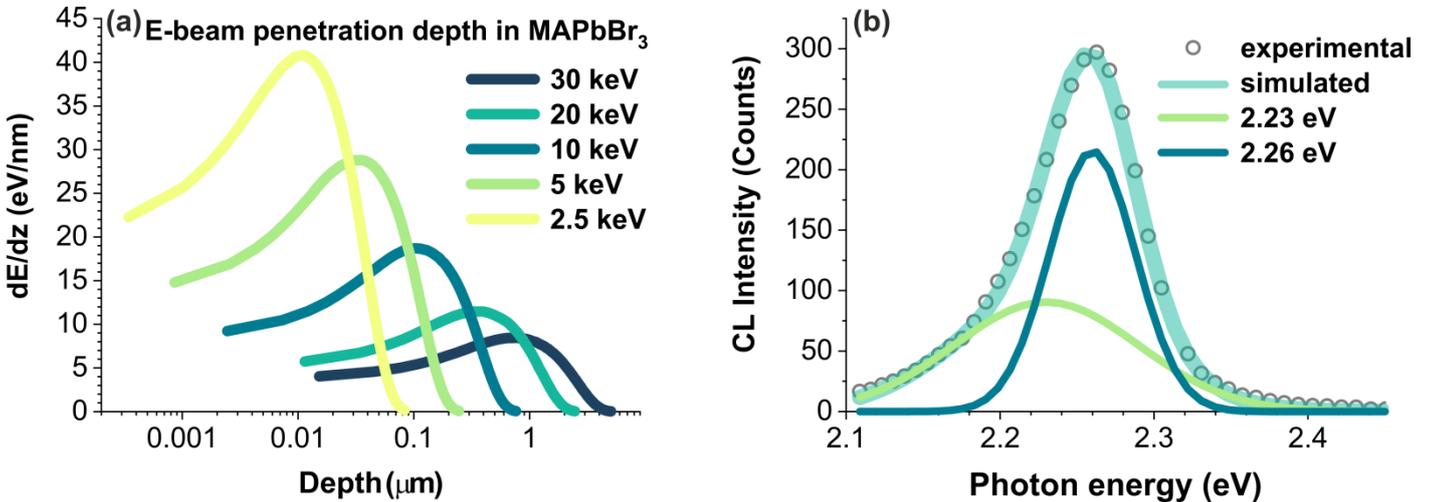

Figure 1 - Depth dependencies of e-beam deposited energy calculated for the 2.5-30 keV e-beam energies (a); CL spectra of "as fabricated" $MAPbBr_3$ MNC measured with the beam energy of 30 keV (300K) and the result of the fitting with two Gaussian peaks (b)

Then, we carried out the CL measurements at $E_b$=30 keV after a set of consequent LEEBI dose increasing up to 43 mC/cm², as shown on the **fig.2(a)**. The obtained results clearly exhibit the significant shifts in the CL peaks with increased energy of the emitted photons up to ~2.55 eV. The intensity of the CL decreased with the raise of the dose compared to the data for "as fabricated" $MAPbBr_3$ MNC. The CL spectra on the irradiation dose can be roughly separate in three regions. At low irradiation doses (up to about 3 mC/cm²) the spectral form practically does not change. However, the luminescence intensity decreases with a dose and the position of maximum gradually shifts to a higher energy (from 2.26 to 2.28 eV). In this region, the spectra can be fitted with three Gaussian peaks with energy of maximum at 2.26 and 2.23 eV and the third band, the energy of which slightly increases with a dose (Fig. S2 in supplement) that corresponds to the "blue" shift. As the dose goes up, the intensity of the 2.26 and 2.23 eV bands reduces, while the intensity of the third band increases (**Fig. 2(b)**).

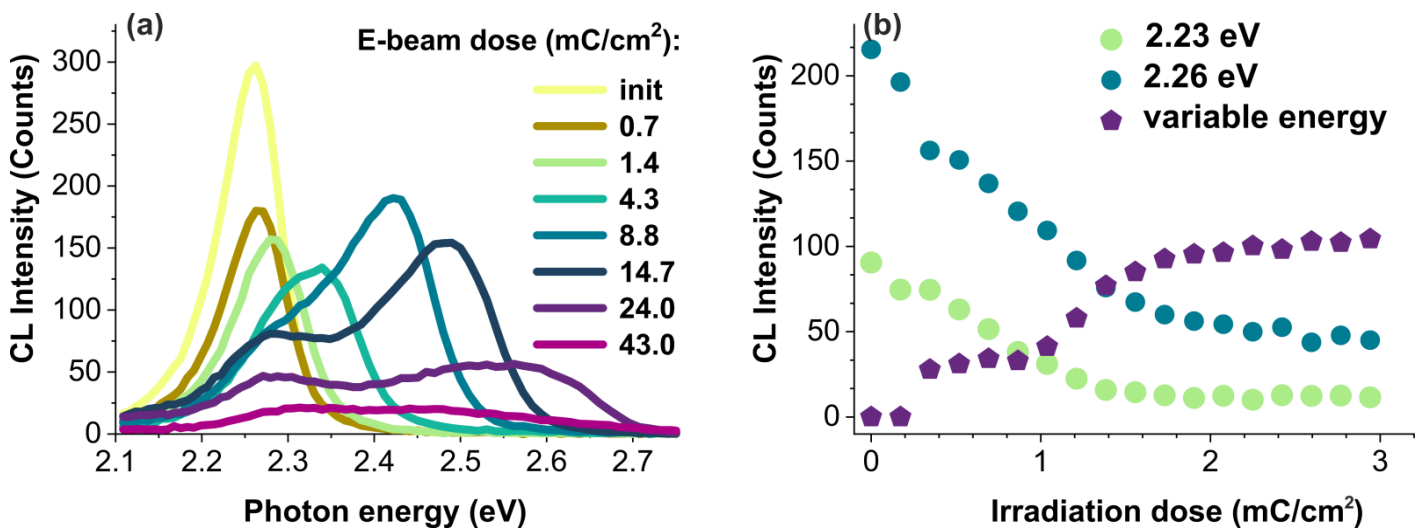

Figure 2 - CL spectrum evolution with irradiation dose at $E_b$= 30 keV and beam current of 1 nA (a); Dose dependencies of CL band intensities for 30 keV irradiation (b)

In the range from 3 to 20 mC/cm$^2$ the spectra drastically change, the shift to higher energy increases, however, the intensity practically does not change and even can increase at some doses (Fig. 2a). The spectra deconvolution shows the complex transformation of optical properties. Thus, the emission band with the energy of 2.38 eV appears, then it is replaced with the 2.41 eV band and then with 2.55 eV one. When the LEEBI dose is over 20 mC/cm$^2$, the CL spectra can be fitted with three Gaussian bands with energies of 2.27, 2.41, and 2.55 eV. The spectra have negligible changes, but its intensity gradually decreases. Such a decrease could be determined by the well-known buildup of hydrocarbon film on the surface after a long exposure to the e-beam[19].

The evolution of CL spectrum under LEEBI with lower beam energies is qualitatively similar. However, at the doses below 3 mC/cm$^2$ the evolution of spectrum occurs faster while beam energy decreases. That allows to assume that the transformation of single crystal MAPbBr$_3$ under LEEBI is not straightly dependent on the beam energy, which only affects the rate of this transformation. As an example, dose dependencies of 2.26 eV band intensity at different beam energies are shown in **Fig. 3(a).** The analysis of the CL intensity vs. irradiation dose shows the maximum slope $D_0$=0.79 for the 5 keV beam energy, while for the $E_b$=30 keV, the $D_0$=0.24. If the decay is approximated with exponential function $\exp(-D_{irr}/D_0)$, $D_0$ increases with beam energy approximately as $E_b^{0.75}$, i.e. inversely proportional to the density of deposited energy from 0.57 at 5 keV to 1.54 at 30 keV. It should be noted that MAPbBr$_3$ MNCs demonstrate the higher stability under LEEBI in comparison with the data reported in the literature[20]. *Yi et al.,* described the drop of the luminescence intensity by about 40% already after 15 µC/cm$^2$ e-beam irradiation with $E_b$= 20 keV. As seen in **Fig. 3**, for samples studied at $E_b$= 20 keV, the intensity lost approximately 50% after the dose of 380 µC/cm². Different beam energies demonstrate the dose dependency of maximum band position with variable energy (as shown in the **fig.S2 in supplementary information (S.I.)**).

The LEEBI effect on the MAPbBr3 MNCs requires consideration of different interaction pathways. Four primary mechanisms of e-beam induced damage are usually considered: atomistic displacement or "knock on" effect [21], heating the sample under exposure caused by thermalization[22]; ionization damage or radiolysis[23], and electrostatic charging[24]. In the experiments carried out in the present work, the beam energy is too low to produce a displacement of atoms. It seems that electrostatic charging could also be neglected because, although the sample resistivity is rather high[25]. No charging effects were observed in the secondary electron mode of the imaging. The heating induced by the beam current is typically negligible, as reported by *Egerton et al.*[26], the temperature increase under focused e-beam excitation can be estimated as $\Delta T = \frac{1.5 I_b E_b}{\pi \kappa R}$, where κ is the thermal conductivity. Taking into account low thermal conductivity of MAPbBr3 (κ = 0.35 Wm$^{-1}$K$^{-1}$ [27]), for the worst-case scenario with $I_b$= 1 nA and $E_b$ = 2.5 keV, the changes in the temperature (ΔT) could be estimated as ~ 50°C, i.e. the sample temperature can rise up to 70-75°C. For $E_b$ equal to 5 and 10 keV ΔT ~ 30° and 19° respectively, for the point source. This value will be even much lower because the beam is scanned over a larger area (the area is of 1.2×10$^4$ µm² for the largest magnification and raises with its decrease). The MAPbBr3 MNC have structural stability up to 85°C[28]. Thus, the critical heating of the sample under LEEBI can be also neglected. The possible thermal decomposition processes, occurring at the room temperature, take place only on the surface[29]. Thus, the most probable mechanism of damage under LEEBI is radiolysis, in which electronic excitation and/or ionization can result in the breaking of bonds (**fig.3b**). The radiolysis effect can occur via recombination-enhanced reaction, in which the energy released during excess carrier recombination is redirected to the local vibration modes[30], or by the Coulomb repulsion of atoms recharged under irradiation. The high-dose irradiation resulted in surface cracking of the MAPbBr3 MNC (**Fig. S3 in S.I.**). We noticed that crack formation strongly depends on $E_b$. The conditions of 10 keV with 25 mC/cm² irradiation were not harmful for the surface of the MAPbBr3 MNC, while at 30 keV cracks appear after a dose of about 5 mC/cm². The similar dependence on $E_b$ was described in [31], where cracking was not observed after 5 keV irradiation but appears after 15 keV irradiation with a dose exceeding 100 mC/cm². That means that the stress rises with the increase in $E_b$, probably due to the increase of damaged layer thickness with $E_b$.

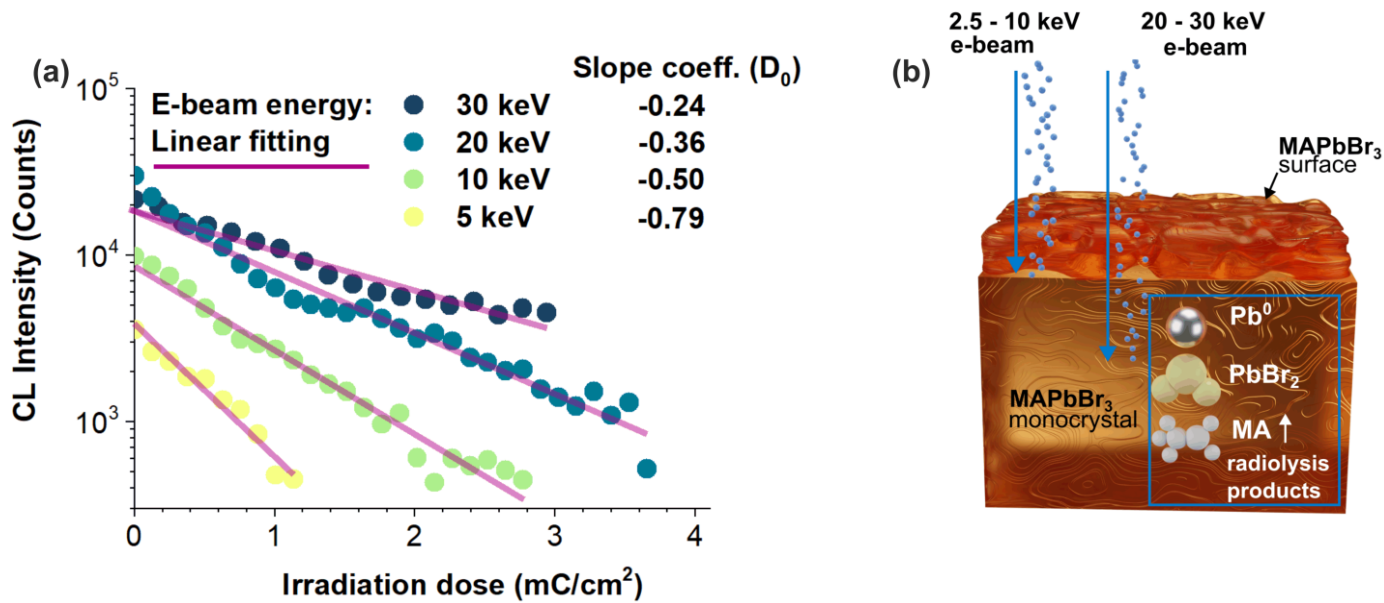

Figure 3 - A decay of 2.26 eV emission band intensity at different $E_b$ (a); the schematics for of the MAPbB

The observed cathodoluminescence peak at 2.26 eV typically associated with the band-to-band transition in single crystal MAPbBr$_3$[32]. The LEEBI experiment showed that the energy of the emission at 2.26 eV decreases with an irradiation dose higher than 3-5 mC/cm$^2$. Higher doses of the electron beam can likely lead to partial decomposition of the perovskite material. As a result of the radiolysis, the gas phases, such e.g. as MA, can be depleted from the surface. Thus, inorganic phases with larger band-gap ($E_{g\ PbBr2}$ = 3.02 eV[33]) may form in the near-surface layer. Cracks formation after high dose irradiation seems to confirm such scenario. Besides, as shown in [34], polycrystalline-like phase can be formed at the surface, which also can lead to the blue shift of emission[27], in particular, it can explain the appearance of the emission band with variable energy. The effect of LEEBI is qualitatively very similar to that previously observed on the perovskite materials, i.e., a noticeable decrease in emission intensity and a blue shift in emission energy[13]. According to our results, the use of 30 keV beam energy provides a more stable CL signal. As the low-energy peak fades, a new emission band emerges, and its intensity is related to the dose of radiation. Such behavior is rather unusual and was not previously reported in the studies of LEEBI effect on the perovskite materials.

We could propose the following interpretation for the evolution of the CL spectrums. The first stage of the degradation under LEEBI is driven by a deviation of the chemical decomposition in MAPbBr$_3$ MNC, that leads to an increase in the band-gap. The deposited energy of the e-beam has the maxima of penetration in the depth of material (accord to the calculations presented on the fig.1). The changes in the chemical composition of MAPbBr$_3$ occur in the volume of an MNC with the highest deposited energy. The changes in the intensity of the 2.26 eV peak and variable emission energy with the raise of the e-beam dose support this assumption.

The 2.23 eV band was measured only at $E_b$ equal to 20 and 30 keV which provides deeper penetration in-depth, compared to lower e-beam energies. The corrosion induced by ambient conditions (contamination with the moisture and oxygen) is spreading from the surface to the volume of the $MAPbBr_3$ MNC, thus, the use of large $E_b>20$ keV eliminates the impact from the degraded material. We found that the decay of the CL intensity has inverse proportionality to the $E_b$ (the rate of the decay was estimated to $E_b^{-0.75}$). The obtained dependence can be originated from reduced energy density or accelerated decomposition processes in the near –surface layers.

Thus, the results obtained clearly demonstrate that the CL measurements can provide the valuable information concerning quasi-chemical reactions in $MAPbBr_3$ under LEEBI. The process of radiolysis should be proportional to the excess carrier concentration, which also straightly correlates to the deposited energy density. Analysis of the CL spectra at different LEEBI showed that the appearance of the emission band with variable energy is almost independent of $E_b$. This contradicts the conclusion made in [35] that reducing acceleration voltage is more favorable for characterization of halide perovskites with the use of e-beam (SEM, CL and related techniques). The presented results have shown that a study of dose and beam energy dependences of LEEBI effect on the $MAPbBr_3$ degradation by the CL method can provide important information elucidating some mechanisms of $MAPbBr_3$ degradation. Our study demonstrates that minimization of the e-beam induced damaged and relative stabilization of the CL intensity for halide perovskite MNCs requires a specific approach. The use of the large e-beam energies (>20 keV) is preferable for eliminating the impact of the surface decomposition and accelerated corrosion processes induced in ambient conditions. Another key point is the decrease in deposition energy density responsible for the radiolysis effects occur in the volume of the material.

See the supplementary material for the detailed experimental section, LEEBI and cathodoluminescence studies, XRD of the $MAPbBr_3$ MNC and SEM images.

**Acknowledgements:**
The work at IMT RAS was supported by the State Task No 075-01304-23-00. Danila Saranin gratefully acknowledges the financial support from the Russian Science Foundation (RSF) with a grant № 21-79-00299.

**DATA AVAILABILITY**

The data that supports the findings of this study are available in this article and its supplementary material.